# SELF-TRAINING WITH NOISY STUDENT MODEL AND SEMI-SUPERVISED LOSS FUNCTION FOR DCASE 2021 CHALLENGE TASK 4

## Technical Report

*Nam Kyun Kim*[1] *and Hong Kook Kim*[1,2]

[1]School of Electrical Engineering and Computer Science, [2]AI Graduate School
Gwangju Institute of Science and Technology
123 Cheomdangwagi-ro, Gwangju 61005, Republic of Korea
{skarbs001, hongkook}@gist.ac.kr

## ABSTRACT

This report proposes a polyphonic sound event detection (SED) method for the DCASE 2021 Challenge Task 4. The proposed SED model consists of two stages: a mean-teacher model for providing target labels regarding weakly labeled or unlabeled data and a self-training-based noisy student model for predicting strong labels for sound events. The mean-teacher model, which is based on the residual convolutional recurrent neural network (RCRNN) for the teacher and student model, is first trained using all the training data from a weakly labeled dataset, an unlabeled dataset, and a strongly labeled synthetic dataset. Then, the trained mean-teacher model predicts the strong label to each of the weakly labeled and unlabeled datasets, which is brought to the noisy student model in the second stage of the proposed SED model. Here, the structure of the noisy student model is identical to the RCRNN-based student model of the mean-teacher model in the first stage. Then, it is self-trained by adding feature noises, such as time-frequency shift, mixup, SpecAugment, and dropout-based model noise. In addition, a semi-supervised loss function is applied to train the noisy student model, which acts as label noise injection. The performance of the proposed SED model is evaluated on the validation set of the DCASE 2021 Challenge Task 4, and then, several ensemble models that combine five-fold validation models with different hyperparameters of the semi-supervised loss function are finally selected as our final models.

## 1. INTRODUCTION

Sound event detection (SED) aims to detect and classify individual sound event categories and their onset and offset in diverse sound environments. SEDs can affect a wide range of applications related to sound sensing [1]. For example, acoustic monitoring can detect physical events, such as glass breakage, gunshots, tire slippage, or car crashes. SED can also be integrated into audio captions [2], audio monitoring in smart cities [3], life support and healthcare [4], etc. to better understand media content.

In general, the SED task requires a large amount of labeled training data, thus hand-labeling these collected data is extremely costly. Moreover, such training data should be collected in a real environment in which a target application using SED could be deployed [1]. As an alternative, limited strongly labeled data are used for model training by combining an ample amount of weakly labeled data whose labels only include the sound event types without any information on the timestamps of the events. Moreover, synthetic audio data could be used to train the model.

The DCASE 2021 Task 4 is the follow-up to DCASE 2020 Task 4. Compared to DCASE 2020 Task 4, this year's task includes an increased amount of strongly labeled data, while weakly labeled data and unlabeled data are identical in both year's tasks. According to the results of the DCASE Challenge 2020 Task 4, some of top-ranked models were based on a mean-teacher model [5] trained by both weakly labeled and unlabeled data with consistency regularization. Specifically, both the teacher and student model in the mean-teacher model [6] were constructed with the same network architecture, and the teacher model aimed to help the student model that was used for SED, where the model parameters of the teacher model were updated by the exponential moving average of the student model parameters.

In this report, we propose an SED model based on the self-training of the student model in the mean-teacher model. In other words, we first construct a residual convolutional recurrent neural network (RCRNN)-based mean-teacher model [7] and then train it using all the training data, including strongly labeled, weakly labeled, and unlabeled data. Then, the trained teacher model is used to predict the strong label for each of the weakly labeled or unlabeled data. Next, a noisy student model, which is initialized by the student model of the mean-teacher model, is learned by using the given labels for strong labeled data and the predicted labels for weakly labeled or unlabeled data. Especially, the self-training approach in [8] is applied to train the noisy student model with noise injection. In particular, we consider three different noise-injection techniques: 1) feature noises, such as SpecAugment [9], mixup [10], and time-frequency shift [11], 2) dropout-based model noise, and 3) a semi-supervised loss function that acts as label noise injection.

Following this introduction, Section 2 summarizes the dataset and explains the pre-processing method used in this work. Then, Section 3 proposes self-training with a noisy student model and semi-supervised loss function, and Section 4 discusses the experimental results on the validation set for DCASE 2021 Task 4. Finally, we conclude this report in Section 5.

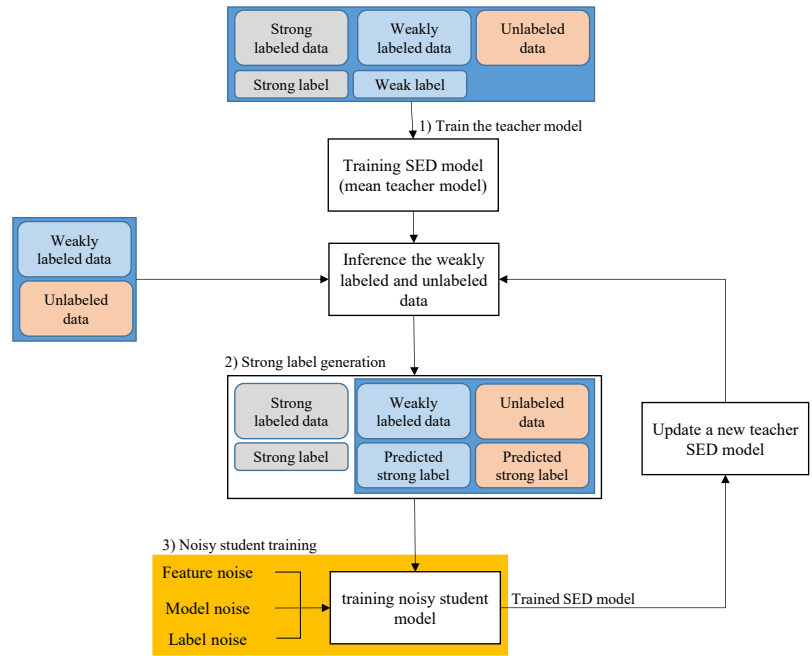

Figure 1: Training procedure of the proposed SED model composed of the RCRNN-based mean-teacher model for predicting strong labels and the self-trained noisy student model with noise injections and a semi-supervised loss function.

## 2. DATASET

The DCASE 2021 Challenge Task 4 consists of three different datasets for model training: 1) a weakly labeled training dataset (without timestamps), 2) an unlabeled in-domain training dataset without any label, and 3) a strongly labeled synthetic dataset. The weakly labeled and unlabeled in-domain training datasets are taken from the AudioSet [12], but the strongly labeled synthetic dataset is generated using the Scaper soundscape synthesis and augmentation library [13]. The weakly labeled training dataset contains 1,578 audio clips with weak annotation only. The unlabeled in-domain training dataset and the strongly labeled dataset contain 14,412 and 10,000 audio clips, respectively. Each audio clip is stored as both mono- and stereo-channel signals that are sampled at 44.1 kHz with a maximum duration of 10 seconds.

For a given dataset, we first take the mono-channel signals and resample them from 44.1–16 kHz. After that, each resampled audio signal is segmented into consecutive frames of 2048 samples with 256 samples of hop length. Then, a 2048-point fast Fourier transform (FFT) is applied to each separated signal, and a 128-dimensional mel-filterbank analysis is performed for each frame. Since each 10-second audio clip is represented by 625 frames, the dimension of the input feature for the SED model is 1×625×128. Note here that zero padding is applied to the audio clips that are shorter than 10 seconds. Finally, the extracted mel-spectrogram features are normalized by the global mean and the standard deviation over all the training audio clips.

## 3. METHOD

Fig. 1 shows the training procedure for the proposed SED model composed of the RCRNN-based mean-teacher model for predicting strong labels and the self-trained noisy student model with noise injections and a semi-supervised loss function. The detailed explanation on the mean-teacher model and noisy student model will be given in the following subsections.

### 3.1. RCRNN-based mean-teacher model

As shown in Fig. 1, the first stage of the proposed SED model is based on an RCRNN-based mean-teacher model proposed in [7], which is also the same architecture in [14] by replacing the CRNN with an RCRNN. Table 1 shows the network architecture and hyperparameters of RCRNN used in the mean-teacher model.

To begin with, the input feature of 625 frames is grouped to make a (625×128) spectral image, which is then used as the input feature for the RCRNN. As described in the table, the convolutional blocks of the RCRNN are composed of one stem block and five residual convolutional blocks, where the stem block consists of two convolutional blocks with 16 and 32 kernels for the first and second convolutional blocks, respectively. Each convolutional block has (3×3) kernels with a stride of (1×1), and it is followed by batch normalization, GLU activation, and a (2×2) average pooling layer. Next, the convolutional block attention module (CBAM)-based attention [15] is applied to the output of each residual convolutional block. After finishing all the residual convolutional blocks, the (128×156×1) feature map is applied to a recurrent block. The recurrent block consists of two bidirectional gated recurrent units (BiGRUs) to learn the temporal context information, where a rectified linear unit (ReLU) is used as an activation function for each GRU. The (156×256) output of the recurrent block is processed by an FC layer and then by a sigmoid function, resulting in a (156×10) output, where 10 denotes the number of sound events to be detected. Note that a (156×10)-dimensional output is related to a strong label including the sound event type and timestamp. Moreover, a weighted pooling layer is applied to the (156×10)-dimensional output to obtain a (1×10)-dimensional output that predicts a weak label for the given audio clip.

Table 1: Network architecture of a residual convolutional neural network in the RCRNN used in the mean-teacher model

| Name | Layers | Output shape |
|---|---|---|
| *Input layer* | Input: log-mel spectrogram | 1×625×128 |
| *Stem block* | (7 × 7, Conv2D, @16, GLU, BN) <br> 2×2 average pooling layer | 16×312×64 |
| | (7 × 7, Conv2D, @32, GLU, BN) <br> 2×2 average pooling layer | 32×156×32 |
| *Residual convolutional block* | $\begin{pmatrix} 3 \times 3, \text{Conv2D}, @64, \\ \text{ReLU, BN} \end{pmatrix} \times 2$ <br> Self-attention module (CBAM) <br> 1×2 average pooling layer | 64×156×16 |
| | $\begin{pmatrix} 3 \times 3, \text{Conv2D}, @128, \\ \text{ReLU, BN} \end{pmatrix} \times 2$ <br> Self-attention module (CBAM) <br> 1×2 pooling layer | 128×156×8 |
| | $\begin{pmatrix} 3 \times 3, \text{Conv2D}, @128, \\ \text{ReLU, BN} \end{pmatrix} \times 2$ <br> Self-attention module (CBAM) <br> 1×2 average pooling layer | 128×156×4 |
| | $\begin{pmatrix} 3 \times 3, \text{Conv2D}, @128, \\ \text{ReLU, BN} \end{pmatrix} \times 2$ <br> Self-attention module (CBAM) <br> 1×2 average pooling layer | 128×156×2 |
| | $\begin{pmatrix} 3 \times 3, \text{Conv2D}, @128, \\ \text{ReLU, BN} \end{pmatrix} \times 2$ <br> Self-attention module (CBAM) <br> 1×2 average pooling layer | 128×156×1 |
| *Recurrent block* | (128 BiGRU cells) ×2 | 256×156 |

The RCRNN-based mean-teacher model trained so far is used to generate predicted strong labels on weakly labeled and unlabeled datasets. Note here that the predicted labels are binary labels by applying a threshold to the sigmoid output, where the threshold is set to 0.5. By using these predicted labels, a noisy student model is trained, which will be described in the next subsection.

### 3.2. Noisy student model

The second stage of the proposed SED model is an RCRNN-based noisy student model whose network architecture is identical to the student model of the mean-teacher model. To train the noisy student model, the strong labels predicted from the mean-teacher model in the first stage in the proposed SED model are used for weakly labeled or unlabeled data, while the given strong labels for strongly labeled data are used. After that, the input spectral image is changed by sequentially applying the noise-injection techniques of time-frequency masking from SpecAugment [9], mixup [10], and time-frequency shift [11]. Time-frequency masking operates by replacing values in the time and frequency domain with zero, and mixup generates a noisy data by mixing current input feature and another one to smooth the distribution of samples in the feature space. Time frequency shifting circularly shifts the input spectral image along the time and frequency axes for a random Gaussian noise with zero mean and a standard deviation of 4 and 32 for the frequency and time axes, respectively. In addition, a dropout with a rate of 0.5 is applied to realize model noise for the noisy student model. Finally, during the training of the noisy student model, a semi-supervised loss function was used for the realization of the noisy target label. The semi-supervised loss function is defined as

$$L_{semi} = \sum_{i \in S} BCE(i;\theta) + \sum_{i \in \{W,U\}} BCE_{soft}(i;\theta) \quad (1)$$

where $S$, $W$, and $U$ indicate the sets of strongly labeled, weakly labeled, and unlabeled data, respectively. In addition, $\theta$ denotes the RCRNN-based noisy student model. In addition, $BCE(i;\theta)$ is binary cross entropy (BCE), and $BCE_{soft}(i;\theta)$ is the BCE between the binarized strong label from RCRNN-based mean-teacher model and the predicted output from $\theta$. In other words, $BCE_{soft}(i;\theta)$ is defined as

$$BCE_{soft}(i;\theta) = -(\bar{y}_i \log \hat{y}_{i,\theta} + (1-\bar{y}_i) \log(1-\hat{y}_{i,\theta})) \quad (2)$$

where $\hat{y}_{i,\theta}$ is the output of the RCRNN-based noisy student model, $\theta$, for the $i$-th audio clip. In (2), $\bar{y}_i$ is an interpolated target between the binarized strong label, and it is computed as

$$\bar{y}_i = \beta \hat{y}_{i,\theta} + (1-\beta)\hat{y}_{i,\theta_m} \quad (3)$$

where $\hat{y}_{i,\theta_m}$ is a binarized strong label of the RCRNN-based mean-teacher model, $\theta_m$. $\beta$ is a hyperparameter for the loss function, and it is set for obtaining differently ensemble models. Consequently, by using the noisy input spectral image, the noisy student model is trained with dropout and the semi-supervised loss function.

After finishing the noisy student model training, the model parameters are copied into the teacher model of the mean-teacher model. Then, the strong labels for weakly labeled or unlabeled data are updated, which are also brought to the noisy student model as new target labels. This procedure for training the noisy student model and updating the labels is repeated once more.

## 4. EXPERIMENTAL RESULTS

### 4.1. Model training

The neural network weights of the mean-teacher model were initialized by using Xavier initialization, but the biases were all initialized to zero. Next, the mini-batchwise adaptive moment estimation (ADAM) optimization algorithm was applied, where dropout was also applied at a rate of 0.5. In addition, the learning rate was set according to the ramp-up strategy, where the maximum learning rate reached 0.001 after 50 epochs. For data augmentation, we employed time-frequency shift [10] and mixup [11]. Meanwhile, the noisy student model was trained as described in Section 3.2 on a basis of 5-fold cross-validations where all the data in the training set were divided into 5 folds, and 4 out of 5 folds were used for training, and the remaining fold was used for validation. Here, the learning rate was initially set to 0.001, and it was reduced by a simple learning rate schedule (commonly known as ReduceLRonPlateau in PyTorch).

Table 2: Comparison of performance metrics of the baseline and different versions of the proposed SED model on the validation set of the DCASE 2021 Challenge Task 4 where the check mark denotes which noise-injection technique is employed.

| Model | Feature noise | Model noise | Label noise | Event-based F1-score | PSDS-scenario 1 | PSDS-scenario 2 |
|---|---|---|---|---|---|---|
| Baseline: CRNN-based mean-teacher model [16] (Single model) | - | - | - | 40.1% | 0.342 | 0.527 |
| RCRNN-based mean-teacher model (single model) | ✓ | ✓ | - | 49.1% | 0.403 | 0.600 |
| RCRNN-based noisy student model (single model) | ✓ | ✓ |   | 51.6% | 0.425 | 0.649 |
|  | ✓ | ✓ | ✓ | 52.3 % | 0.434 | 0.659 |
| RCRNN-based noisy student model (5-model ensemble) | ✓ | ✓ |   | 53.6% | 0.449 | 0.675 |
|  | ✓ | ✓ | ✓ | 53.9% | 0.450 | 0.682 |
| RCRNN-based noisy student model (Top1-5 ensemble) | ✓ | ✓ | ✓ | 54.4% | 0.451 | 0.679 |
| RCRNN-based noisy student model (Top1-10 ensemble) | ✓ | ✓ | ✓ | 55.4% | 0.457 | 0.685 |

## 4.2. Discussion

The performance of the proposed SED models was evaluated by using the measures defined in the DCASE 2021 Challenge Task 4 [16], which were an event-based F1-score and a polyphonic sound detection score (PSDS) [17]. The PSDS was computed as two different scenarios by using a public open-source software where the parameters of ($\rho_{DTC}, \rho_{GTC}, \rho_{CTTC}, \alpha_{CT}, \alpha_{ST}$) were set to (0.7, 0.7, 0.0, 0.0, 1.0) for PSDS scenario 1 and (0.1, 0.1, 0.3, 0.5, 1.0) for PSDS scenario 2, which were all followed by the rule defined in [16]. Notice that the validation set was composed of 1,168 audio clips with strong labels including time-stamps.

Table 2 compares the performance between the baseline and different versions of the proposed SED models on the validation set of DCASE 2021 Challenge Task 4. The different versions of the proposed SED model included 1) an RCRNN-based mean-teacher model that was the first stage of the proposed SED model, 2) an RCRNN-based noisy student model that was trained by using all the training data without any cross-validation, and 3) three different ensemble models. To construct the ensemble models, each 5-fold model was trained using five different settings of the hyperparameter, $\beta$, of the semi-supervised loss function in Eq. (3) from 0.3 to 0.9 at a step of 0.2, which resulted in five different models for each fold. Then, the model that had the highest F1-score among five different models was selected as the best model corresponding to this fold. After that, the 5-fold models were linearly combined to form an ensemble classifier, which was denoted as a 5-model ensemble, as shown in the fourth row of Table 2. Instead of first selecting the best model for each fold according to different $\beta$s, we selected the top-ranked models from 25 models that were all the models from five different folds and five different $\beta$s. Based on this approach, we constructed two ensemble models, such as the Top1-5 ensemble and Top1-10 ensemble, which are shown in the fifth and sixth row of Table 2, respectively.

As shown in Table 2, the RCRNN-based mean-teacher model achieved a higher F1-score, PSDS-scenario 1, and PSDS-scenario 2 by 9.0%, 0.061, and 0.073, respectively, than the baseline. Moreover, the proposed noisy student model (single model) with feature, model, and label noise injections achieved further improvements of 3.2%, 0.031, and 0.059 for F1-score, PSDS-scenario 1, and PSDS-scenario 2, respectively, compared to the RCRNN-based mean-teacher model. Importantly, the ensemble model of the proposed noisy student model, which was constructed by the Top1-10 ensemble, provided the highest F1-score, PSDS-scenario 1, and PSDS-scenario 2 among all the possible ensemble models.

## 5. CONCLUSION

This report proposed a polyphonic SED model for the DCASE 2021 Challenge Task 4. The proposed SED model was based on self-training with a noisy student model to deal with the different combinations of training datasets, such as the weakly labeled, unlabeled, and strongly labeled synthetic datasets. Especially, the target label of each audio clip from weakly labeled or unlabeled datasets was predicted using the RCRNN-based mean-teacher model. To realize self-training, data augmentation-based feature noise, dropout-based model noise, and semi-supervised loss function-based label noise were injected into the noisy student model. Especially, the noisy student model was trained according to cross-validations and different hyperparameter values of the semi-supervised loss function, which resulted in different ensemble models. The performance of different versions of the proposed SED model was evaluated on the validation set of the DCASE 2021 Challenge Task 4. Consequently, it was shown that the Top1-10 ensemble model improved the F1-score, PSDS-scenario 1, and PSDS-scenario 2 by 15.3%, 0.115, 0.158, respectively, compared to the baseline.

## 6. ACKNOWLEDGMENT

This work was partly supported by Institute of Information & communications Technology Planning & Evaluation (IITP) grant funded by the Korea government(MSIT) (No.2021-0-00014, Development of Audiovisual Cognitive Intelligence Solution based on Edge Computing for Respond to Disaster Situations) and IITP grant funded by the Korea government(MSIT) (2019-0-00447, Development of emotional expression service to support hearing/visually impaired)